# On presentation of lightning flash in thundercloud as collective spontaneous radiation (superradiation) pulse.

M.D. Scherbin

*Federal State Institution*
*"Russian Federation Ministry of Defense*
*12 Central Scientific Research Institute"*

---

∗ Corresponding Author Address: Dr. M.D. Scherbin, Federal State Institution "Russian Federation Ministry of Defense, 12 Central Scientific Research Institute", 141307, Sergiev Posad. E-mail: mdsh@spnet.ru

# ABSTRACT

In this paper we conjecture that some specific properties of an electroactive region of thunderclouds play the major role in an initiation of a lightning flash. The electroactive region is modeled here as a quantum two-level system of oxygen molecules which are assumed to have two (ground and excited) states, where transitions from ground to excited states come as a result of micro-discharges between ice particles. In this case lightning flash corresponds to the pulse of collective spontaneous radiation caused by the deexcitations of excited oxygen molecules from metastable state $^1\Delta_g$ into ground state.

**Keywords**: *lightning, thunderstorms, superradiation, self-focusing, lasers.*

**Index Terms**: *3324 Atmospheric Processes: Lightning; 3304 Atmospheric Processes: Atmospheric electricity*

# Introduction

In our previous paper [1] for us to explain the observed in the photo-images given in [2] the twisting of a lightning channel we supposed that there exists the longitudinal magnetic field in end areas of a bipolar lightning. The estimations of intensities of magnetic and induced (by magnetic) electric fields gave their extreme values, which provide the possibility of the birth of electron-positron pairs in the leader heads.

The existence of such fields allowed us in [1] to explain some particularities of the developing of a lightning flash in a thundercloud:

- the stepwise shape of a negative leader (stepped negative leader)
- anomalously big currents in the positive leader
- more intensive RF (radio-frequency) radiation from the negative leader than from the positive one
- γ-radiation from thunderclouds
- and others

Also, the step shape of a negative leader, as well as the lightly developed step shape of a positive leader, was explained in [1] as appearing due to ponderomotive forces in the vicinities of the leader heads. The difference in the character of the leader propagations was associated to the fact of a greater mobility of electrons in comparison to ions. The physical nature of the extreme intensities of electric and magnetic fields was not revealed in [1]. The possible origin of such intensities will be suggested below.

## Basic suggestions

The possibility of generation of electromagnetic fields of such extreme intensities in a gas medium with high density of energy in local space regions (a leader head of a lightning channel has a size scaled as a few centimeters while a screening zone amounts to the scale of several meters) is objectively can be related to some physical processes taking place in a thundercloud. These processes provide pumping of energy into local areas and its release in a short time period which is characteristic time interval for a lightning flash development. It appears that these processes must be similar to those processes taking place in gas lasers.

It is necessary to find out the following: what could be the active medium in a thundercloud; what are mechanisms of energy pumping; what is the possible way of a releasing the accumulated energy in an active medium.

The best candidate for the role of active medium in a thundercloud, in our opinion, might be the oxygen $O_2$ molecules with first excited metastable state $^1\Delta_g$. Excitation energy is approximately ~ 1eV. Numerous papers are devoted to analysis of this excited state of an oxygen molecule in the atmosphere. It was Lowke who probably was the first to pay attention to the necessity of accounting this excited state in lightning processes (we refer the reader to his papers on developing of lightning channel formation model [3-5]). Excited states $^1\Delta_g$ of a oxygen molecule are developed at different electro-discharge types [6-12]. They may account for 10-20% of all the molecules $O_2$ in the discharge area [6,9,12]. The lifetime $T_1$ of $^1\Delta_g$ state under normal conditions amounts to 45 minutes [7]. In the case of collision processes in the perturbed medium it drops down to 100 ms [3, 10, 11]. To accumulate energy in local thundercloud areas there must be conditions for creating sufficient population inversion within this lifetime interval. Prerequisites for the possibility of such conditions in a thundercloud are available.

In some papers, for example [13, 14], it is pointed out that prerequisites of lightning flash initiation must be locally high concentration, higher than average values, of ice particles. The micro-discharges between them (streamer processes) in electric field must cause the charge increase on ice particles and the local field amplification, which can probably generate the ionization wave, contributing to appearing of lightning leader process development. Local concentrations of ice particle local concentrations can achieve the values of 1000 particles per liter [16].

In papers [15, 16] it is pointed out that ice particles in a thundercloud, which are columns and plates of various shapes, are aligned along the power lines of electric field and assume chain-type structure.

For the purpose of present work the results presented in the paper [13] are crucial. This paper presents the laboratory investigation of a development of positive streamer discharges between ice particles in an external electric field. It is shown that with an increase of electric field intensity over 300 kV/m the multiple process of micro-discharges between particles is starting. This effect is due to (together with the increase of an intensity of electric field) the particle number increase because of their destruction into fragments caused by micro-discharges.

As mentioned above, during micro-discharges there are excited states $^1\Delta_g$ developing. With growing of number of micro-discharges there should be an increase in the population inversion of excited states of oxygen molecules. In paper [13] it is pointed out that in nature the volume of electroactive zones in thunderclouds can reach 10 m$^3$.

The problem is to determine the critical level of population of the excited states $^1\Delta_g$ and to determine the reasons of deexcitations, that is the energy release. Taking into account the large spacial extent of electroactive zone with molecule $O_2$, the transition from $^1\Delta_g$ state to the ground one is possible due to collective spontaneous radiation (superradiation). The superradiation mechanism was for the first time found by Dicke [17] and presented in detail in [18]. The following description comes mainly from [18].

Dicke showed that the system of N inverted two-level atoms (molecules) can spontaneously make a transition from the excited state to the ground state at a time inversely proportional to the number of atoms (molecules) $\tau_c \sim 1/N$. This effect is due to induced correlation between moments of transitions of spatially separated radiators interacting with each other through the radiation field. As a result, atoms (molecules) being in macroscopically big volume are emitting coherently. As total energy W, irradiated by the ensemble of N atoms (molecules), is equal to

$$W = Nh\omega_o, \quad (1)$$

where $\omega_o$ - transitional frequency, then radiation power will be

$$P \approx Nh\omega_o/\tau_c \sim N^2 \quad (2)$$

Therefore, the peak intensity of collective spontaneous radiation is several orders higher than the intensity of spontaneous (incoherent) radiation proportional to N. Dicke superradiation from macroscopicaly large, elongated and open on both ends cylinder of length L and base area A is carried out to small spacial angles to both sides along the largest volume stretch.

Superradiarion is of collective type when populations of upper and lower energy levels of active material are equal to each other (Dicke state). For the considered case with active material presented by oxygen molecules it can be concluded that oxygen molecules in the excited states $^1\Delta_g$ should be ~ 50% of total oxygen content in electroactive thundercloud zone which is possible at the increase of micro-discharge number in the active thundercloud medium while electric field intensity is growing.

We should estimate the possible radiation power from the thundercloud electroactive area and $\tau_c$ pulse duration which are resulting from relations presented in [18]. We assume the normal atmosphere state; dependence of concentration of $O_2$ on the height is not taken into account because only orders of magnitude are of interest.

$$\tau_c = \frac{4\pi A}{\lambda^2} \cdot \frac{T_1}{N} \quad (3)$$

where $\lambda$ - radiation wave length, equal to $1,24 \cdot 10^{-6}$ m.

Electroactive zone volume is taken as 10m³ and it is considered as a cylinder of 1m² base area A. The number of N oxygen molecules in an electroactive zone is adopted as equal to ~ $5,6 \cdot 10^{25}$. As a result, superradiation pulse duration $\tau_c$ is equal to ~ 15 fc.

We calculate the radiation power from the relation (2). At above mentioned parameter values it is equal to ~ $6 \cdot 10^{20}$ Wt.

Due to a spacial inhomogeniety of a medium and due to the effect of self-acting in strong light waves, conditioned by the dependence of complex dielectric permittivity on the wave intensity, there is the possibility of the self-focusing of radiation from a thundercloud electroactive zone when its power exceeds a certain critical value. The problems of nonlinear optics connected with radiation self-focusing are fully enough considered in scientific and review papers, for example in [19, 20]. The critical power value $P_{cr}$ let us evaluate from the relation presented in [21].

$$P_{cr} = \lambda^2 / 2\pi n_0 n_2 , \qquad (4)$$

where $n_o = 1$ - index of air refraction; $n_2 = 5 \cdot 10^{-23}$ m$^2$/Wt - coefficient of nonlinear air medium susceptibility [21, 22].

The value of $P_{cr}$ at the found values of parameters is equal to ~ $5 \cdot 10^9$ Wt. In this case the radiation from a thundercloud electroactive zone must focus at some distance from it.

We estimate the value of $E$ – electric field intensity in self-focusing ray from the known relation

$$P = \frac{c^2 A E^2}{4\pi} \qquad (5)$$

where $c$ – speed of light.

It follows from equation (5) that $E \approx 3 \cdot 10^{11}$ V/m, that is $E$ value is close to that obtained in [1] from other assumptions. In the region of focus the value of $E$ can be such as to provide the formation of electron-positron pairs.

## Discussion

Physics of propagation of strong short-pulse radiation from laser systems, including propagation in air, is being intensively studied nowadays. Nonlinear effects associated with ray propagation in hot and cold plasma and account of relativistic corrections (for example [21, 23-28]) have been thoroughly considered.

It is shown, that due to these effects, when exceeding a certain threshold radiation power, the electron cavitation is developed – that is extraction from plasma medium by arising ponderomotive forces and carrying them away by electromagnetic wave. In principle, the role of ponderomotive forces in propagation of laser radiation under considered conditions is qualitatively analogous to the role of such forces near a lightning head, considered earlier in [1], where to explain its tearing off we used general classical consideration of current-field interaction. The essence of this effect reasonably follows from the investigations presented in [21, 23-28].

It was mentioned above that possible superradiation from a thundercloud electroactive zone occurs at similar intensity in two opposite directions. It is obvious that the polarity of the succeedently developing lightning channels is determined by microdischarge polarity in the electroactive zone. Besides the superradiation probably determines the first stage formation of bipolar lightning leaders.

It is necessary to stress attention to the following.

During self-focusing from thundercloud electroactive area in the vicinity of the focal point, as mentioned above, the electric field intensity can increase up to extreme values, which provide the birth of electron-positron pairs. Energy density is sufficient for total air medium ionization in the vicinity of focal point. In this respect the investigation of laser radiation interaction with targets (for example, in [29-31]) is of a great interest. Upon interaction of a laser radiation with developing plasma causes spontaneous growth of axial magnetic field. The existence of such field in the vicinity of lightning leader head was assumed in [1] to explain the observed twisting of lightning channel. Moreover, in [32, 33] it is shown that when coherent strong radiation from lasers is propagating there should occur spin moments alignment of electrons along ray propagation in the media, due to the processes of multiphoton ionization. What was also hypothetically assumed in [1] to explain the peculiarities of lightning leader propagation.

The formation of the lightning flash during subsequent stages should be determined by high electromagnetic energy density in the lightning leader heads and by periodic pumping of energy from electric to magnetic, as it is stated in paper [1].

**Conclusion**

The presented analysis of the problem together with the suggestions presented in paper [1] provide the basis to believe that a thundercloud can be considered as an irradiating high power system where oxygen molecules are an active medium. The extreme values of electric and magnetic field intensities are concentrated in rather small spacial-temporal intervals. By their nature they are consequences of quantum processes and are not large-scale, which should impede their registration. Though their screening by surrounding lightning channel layers of charged air medium is possible.

Rather reasonable objectivation of above presented ideas on lightning flash initiation in thundercloud can be the confirmation of the increase of population in oxygen molecules excited state $^1\Delta_g$ when microdischarge number is increasing in air medium between ice particles in external electric field. The tendencies of this phenomenon may probably be fixed in laboratory experiments similar to those stated in [13]. Nevertheless, the results of nature observations may indirectly confirm the existence of electroactive zones in a thundercloud presenting the two-level irradiating system. In [35] they refer to the paper [35], where it is observed luminescence in thunderclouds during 100 ms before the appearing of negative leaders. In paper [36] there has been expressed a supposition that a certain "predischarge" inside the cloud proceeds the appearing of negative leaders. It is more reasonable to suppose that luminescence occurs due to the formation of two-level system and working out by radiation the collective rate of the following superradiation pulse.